\documentclass[epj]{svjour}
\usepackage{graphicx}
\usepackage{bm}
\begin{document}

\title{Ion-induced electron production in tissue-like media and DNA damage mechanisms}
\author{E. Surdutovich\inst{1, 2}\thanks{\emph{E-mail:surdutov@oakland.edu; Tel:+1-248-370-3409}} \and O. I. Obolensky\inst{1,5} \and E. Scifoni\inst{1} \and I. Pshenichnov\inst{1,3}\and I. Mishustin\inst{1,4} \and A. V. Solov'yov\inst{1,5}\thanks{\emph{E-mail:solovyov@fias.uni-frankfurt.de}} \and W. Greiner\inst{1}
}                     

\institute{Frankfurt Institute for Advanced Studies,
Ruth-Moufang-Str. 1, 60438 Frankfurt am Main, Germany \and Department
of Physics, Oakland University, Rochester, Michigan 48309, USA \and
 Institute for Nuclear Research, Russian Academy of Science, 117312
Moscow, Russia \and Kurchatov Institute, Russian Research Center, 123182 Moscow, Russia \and \emph{On leave from A.F. Ioffe Physical-Technical Institute, 194021 St. Petersburg, Russia}}

\date{Received: March 3, 2008 / Revised version: date}
%
\abstract{
We propose an inclusive approach for calculating characteristics of secondary electrons produced by ions/protons in tissue-like media.  This approach is based on an analysis
of the projectile's interaction with the medium on the microscopic
level. It allows us to obtain the energy spectrum and abundance of secondary  electrons
 as functions of the projectile kinetic energy. The physical information obtained in this analysis is
related to biological processes responsible for the irrepearable DNA damage induced by the projectile. In
particular, we consider double strand breaks of DNA caused by
secondary electrons and free radicals, and local heating in the ion's
track. The heating may enhance the biological effectiveness of electron/free
radical interactions with the DNA and may even be considered as an
independent mechanism of DNA damage.  Numerical estimates are
performed for the case of carbon-ion beams.  The obtained dose-depth curves
are compared with results of the MCHIT model based on the GEANT4
toolkit.
\PACS{
      {61.80.-x}{Physical radiation effects, radiation damage} \and
      {87.53.-j}{Effects of ionizing radiation on biological systems}   \and
      {41.75.Aki}{Positive-ion beams} \and
      {34.50.Bw}{Energy loss and stopping power}
     } 
} 
\maketitle
\section{Introduction}
\label{intro}
Ion and proton beams are becoming more and more popular tools for cancer therapy. Ions and protons are more
advantageous projectiles than the now conventional photons because they may
cause less damage to the healthy tissues surrounding tumors and thus induce
fewer side effects. More details on the history, current status and
comparison of proton and heavy-ion therapies can be found in the
review articles~\cite{Kraft05,Pedroni00,Pedroni05,Smith06}.

The method of radiation therapy consists in the inactivation of cancer cells via radiation impact on their DNA~\cite{Goodhead06}. The DNA 
may be destroyed in a number of ways~\cite{Nikjoo99}. Double strand
breaking (DSB) 
is one of the most effective mechanisms of irreparable DNA
damage. Interactions of low-energy secondary electrons and free
radicals with the DNA are believed to be mainly responsible for the
DSB's~\cite{Nikjoo06}. We are going to build an inclusive approach yielding such microscopic characteristics as abundance of
secondary electrons and free radicals in the region as well as their energy
spectrum. These characteristics can be obtained from the analysis
of the interaction of projectiles with the medium.  We hope that this
approach will facilitate establishing a quantitative connection
between the amount of energy deposited into the tissue and the induced
biological consequences.  Another mechanism that we
discuss in this paper is the DNA damage due to local heating of the
medium.

Regardless of the eventual mechanism of the DNA damage, a projectile's
propagation and stopping is the basic process in cancer therapy and therefore we
start our study with its analysis.  All secondary electrons and free
radicals that may cause DNA damage as well as the medium's temperature
increase are due to energy loss by these projectiles.  For numerical
estimates, we have taken parameters characteristic of cancer radiation
therapy using carbon-ion beams.

When an energetic charged projectile enters a medium, it experiences a
number of atomic and nuclear interactions. Nuclear processes, mainly
fragmentation reactions, result in the transformation of the beam
particles into new species. The nuclear processes can be reliably
modeled with the use of Monte-Carlo simulations, in particular by
means of the MCHIT model~\cite{Mishustin05,Mishustin06,Mishustin07}
based on the GEANT4 toolkit~\cite{Agostinelli03,Allison06}.  Therefore
in this paper, we only comment on projectile fragmentation processes
without considering them in detail. Atomic processes, such as
elastic Coulomb scattering, scattering with excitation, single and
multiple ionization, are much more probable but less destructive
events for the projectile, frequently occurring along the projectile's trajectory.
Ionization of molecules of the target media is the main channel of the
projectile's energy loss. This process is the most important for
biological applications since it results in a shower of secondary
electrons, which may inflict more DNA damage than the projectile
itself.

In Section~\ref{ionpass}, we study the passage of a single ion through the
medium, analyze the impact ionization cross section, linear energy transfer (LET), and evaluate
the abundance of secondary electrons and their energy
spectrum. Calculations are carried out for parameters typical for
cancer therapy. They are compared with experimental results and the
predictions of the MCHIT model for the depth-dose distributions. In Section~\ref{estimates}, we
estimate the densities of electrons and free radicals
produced by the projectiles. Finally, we evaluate the local
heating associated with the energy loss, and estimate DNA damage due to thermal effects.

\section{Ion passage through the medium}
\label{ionpass}

We first describe the passage of a single ion through biological
medium. The medium is approximated by liquid water, which, anyhow, is the
major component of most living tissues.  The ion typically enters
the medium with a sub-relativistic velocity (the corresponding energy
range used in carbon-ion therapy is 100--400 MeV per nucleon) and then
gradually slows down because of energy loss in inelastic
processes. As mentioned above, the impact ionization is the dominant
energy-loss process. Most importantly, this process results in the
production of secondary electrons, which are directly or indirectly
responsible for the radiation damage of DNA.

The inelastic cross sections are smaller at the beginning of the
trajectory (at high velocities) but then increase and reach a maximum
at the so-called Bragg peak. In the vicinity of this maximum, the ion
looses its energy at the highest rate. As its velocity
decreases further, the cross section drops because of kinematic
constraints.

In our first analysis~\cite{ourNIMB}, we have limited our
consideration to non-relativistic projectile velocities.  However, the
relativistic effects are significant for 400~MeV/u ions, and in this
work, we include relativistic corrections. As shown below,
this results in better agreement of our calculations with the measured
position of the Bragg peak.

Emission of electrons (frequently called $\delta$-electrons) in
proton collisions with atoms and molecules has been under theoretical and
experimental investigation for decades~\cite{Rudd85,Rudd92}.  The
quantity of interest is the probability to produce $N$ secondary
electrons with kinetic energy $W$, in the interval $d W$, emitted
from a segment $\Delta x$ of the trajectory of a single ion at the
depth $x$ corresponding to the kinetic energy of the ion, $T$, within
the solid angle $d\Omega$. This quantity is proportional to the
doubly-differentiated cross section (DDCS)
\begin{eqnarray}
 \frac{d^2N(W,T)}{dW d\Omega}= n\Delta x \frac{d^2\sigma (W,T)}{dW
 d\Omega}~,
\label{eq1}
\end{eqnarray}
where $n$ is the number density of water molecules (at standard
conditions $n \approx 3.3\times 10^{22} {\rm cm}^{-3}$).

In our current analysis, the angular distribution of emitted electrons
is not important, therefore we integrate over the solid angle and the
DDCS is replaced with a single-differential cross section
(SDCS). Thus, the total impact ionization cross section,
differentiated in secondary electron kinetic energy, $d\sigma(W,
T)/dW$, becomes the main quantity in our analysis.  Besides the
kinetic energy of secondary electrons and the properties of water
molecules, the SDCS depends on the kinetic energy of the projectile
$T$ and its charge $z$.  We use the semi-empirical Rudd's
expression~\cite{Rudd92} for the SDCS, which is a parametric
adjustment that combines the experimental data, calculations within
the plane wave Born approximation and other theoretical models.  It is
given in the following form~\cite{Rudd92}:
\begin{eqnarray}
\frac{d\sigma (W,T)}{dW}= z^2 \sum\limits_i \frac{4\pi a_0 N_i}{I_i}
\left(\frac{R}{I_i}\right)^2 \times \\ \nonumber
\frac{F_1(v_i)+F_2(v_i)\omega_i}{\left(1+\omega_i\right)^3\left(1+\exp(\alpha
(\omega_i-\omega^{\rm max}_i)/v_i)\right)}~,
\label{sdcs}
\end{eqnarray}
\noindent where the sum is taken over the electron shells of the water
molecule, $a_0 = 0.0529$~nm is the Bohr radius, $R=13.6$ eV is the
Rydberg, $N_i$ is the shell occupancy, $I_i$ is the ionization potential of
the shell, $\omega_i=W/I_i$ is the dimensionless normalized kinetic
energy of the ejected electron, $v_i$ is the dimensionless normalized projectile
velocity given by
\begin{equation}
v_i=\sqrt{\frac{m V^2}{2 I_i}}~,
\label{vi}
\end{equation} 
\noindent where $m$ is the mass of electron and $V$ is the velocity of
the projectile. In the non-relativistic case, the
expression~(\ref{vi}) is the square root of the kinetic energy of an
electron having the same velocity as the projectile, normalized by the
corresponding binding energy. The definition of $v_i$ remains in the
relativistic case, however, the projectile velocity $V$ is no longer equal to
$\sqrt{\frac{2T}{M}}$ (where $M$ is the mass of a projectile), but is
rather given by $\beta c$, where
$\beta^2=1-1/\gamma^2=1-(Mc^2/(Mc^2+T))^2$, and $\gamma$ is the Lorentz
factor of the projectile.

\begin{figure*}[th]
\resizebox{1.85\columnwidth}{!}{
\includegraphics{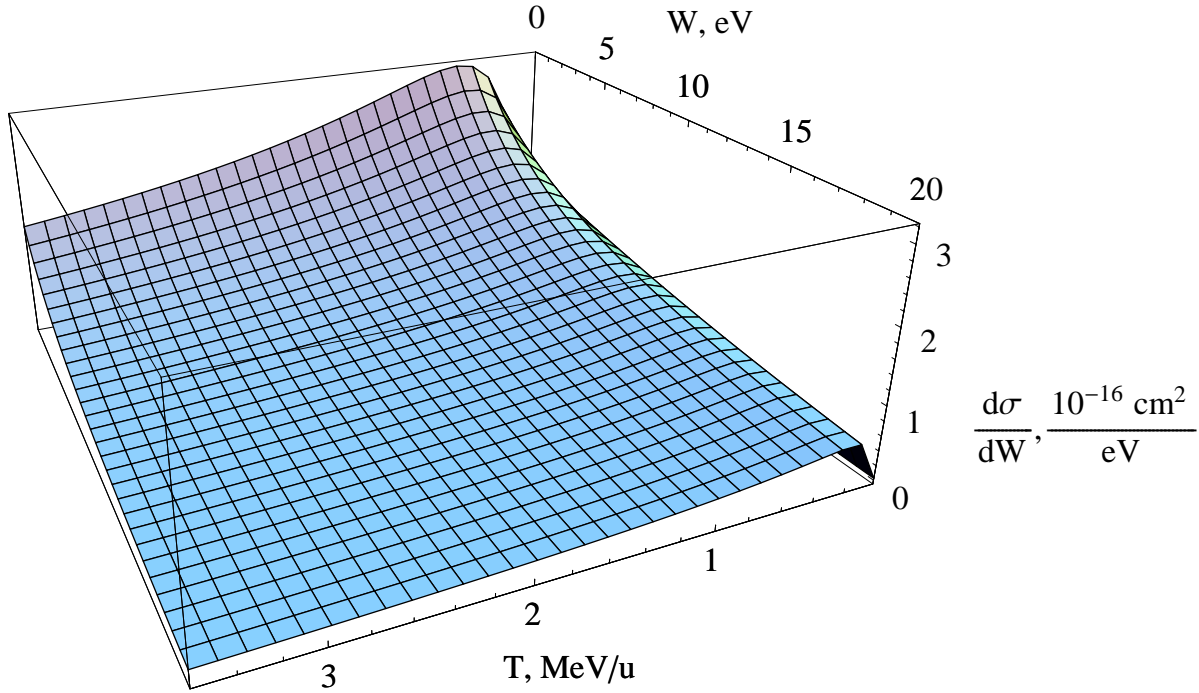}
}
\caption{\label{cs3d} Singly differentiated ionization cross section
for ions $^{12}$C$^{6+}$ interacting with water as a function of the
projectile's kinetic energy $T$ and the energy of ejected electrons
$W$.}
\end{figure*}

Furthermore, $F_1$ and $F_2$ in (\ref{sdcs}) are given by
\begin{equation}
F_1(v) = A_1 \frac{\ln(1+v^2)}{B_1/v^2+v^2} + \frac{C_1 v^{D_1}}{1+E_1
v^{D_1+4}}~,
\label{F1}
\end{equation}
and
\begin{equation}
F_2(v) = C_2 v^{D_2} \frac{A_2 v^2+B_2}{C_2 v^{D_2+4}+A_2 v^2+B_2}~.
\label{F2}
\end{equation}
The corresponding fitting parameters taken from
Ref.~\cite{Rudd92}, $A_1$ ... $E_1$, $A_2$ ... $D_2$, $\alpha$ are
listed in tables I and II.
\begin{table}
\caption{Fitting parameters for the three outer shells of water
molecule with the ionization potentials $I_1 = 12.61$~eV, $I_2 =
14.73$~eV, $I_3 = 18.55$~eV.}
{\begin{tabular}{cccccccccc}
\hline\noalign{\smallskip}
$A_1$ & $B_1$ & $C_1$ & $D_1$ & $E_1$ & $A_2$ & $B_2$ & $C_2$ & $D_2$ &
$\alpha$ \\
\noalign{\smallskip}\hline\noalign{\smallskip}
0.97 & 82 & 0.4 & -0.3 & 0.38 & 1.04 & 17.3 & 0.76
& 0.04 & 0.64\\
\noalign{\smallskip}\hline
\end{tabular}}
\label{outer}
\end{table}
\begin{table}
\caption{Fitting parameters for two inner shells of water molecule
with the ionization potentials $I_4 = 32.2$~eV, $I_5 = 539.7$~eV}
{\begin{tabular}{cccccccccc}
\hline\noalign{\smallskip}
 $A_1$ & $B_1$ & $C_1$ & $D_1$ & $E_1$ & $A_2$ & $B_2$ & $C_2$ & $D_2$ & $\alpha$ \\
\noalign{\smallskip}\hline\noalign{\smallskip}
1.25 & 0.5 & 1.0 & 1.0 & 3.0 & 1.1 & 1.3 & 1.0 & 0.0 & 0.66\\
\noalign{\smallskip}\hline
\end{tabular}}
\label{inner}
\end{table}
\noindent The cut-off energy $\omega^{\rm max}$ is given by
\begin{equation}
\omega^{\rm max}_i = 4v_i^2-2v_i-\frac{R}{4I_i}~,
\end{equation}
\noindent where the first term on the right hand side represents the
free-electron limit, the second term represents a correction due to
electron binding, and the third term gives the correct dependence of
the SDCS for $v\ll 1$ \cite{Rudd92}. For $v\gg 1$, Rudd's formula
would have the correct relativistic asymptotics if $F_1$, given
by~(\ref{F1}), is replaced by the following expression,
\begin{equation}
F_1(v) = A_1 \frac{\ln(\frac{1+v^2}{1-\beta^2})-\beta^2}{B_1/v^2+v^2}
+ \frac{C_1 v^{D_1}}{1+E_1 v^{D_1+4}}
\label{F1rel}
\end{equation}
\noindent whose asymptotic behaviour is clearly similar to the well known Bethe-Bloch formula for energy loss:
 \begin{equation}
-\frac{dT}{dx}\sim\frac{1}{\beta^2}\left[\ln\left(\frac{2mc^{2}\beta^{2}}{\left\langle I\right\rangle (1-\beta^{2})}\right)-\beta^{2}\right]~,
\label{BB}
\end{equation}
\noindent
where $\left\langle I\right\rangle$ is the average ionization potential for water molecule.
This correction reveals itself as an increase of the cross section at
high energies. In our calculations, we omitted the term  $-\beta^{2}$ in the numerator of Eq.~(\ref{F1rel}) after checking its negligible contribution in the energy range of interest.
 The SDCS for water as a function of the 
projectile's and secondary electron's energies is plotted in
Fig.~\ref{cs3d}.

In our previous consideration~\cite{ourNIMB}, the charge of the
projectile, $z$, which enters into the expression (\ref{sdcs}) was taken
to be constant and equal to the charge of a fully stripped projectile
nucleus.  This led to a prediction of unphysically large heights of
the Bragg peak in disagreement with experimental data.  This
indicates that the charge transfer, or electron capture by a charged
projectile, takes place in the vicinity of the Bragg peak, effectively
reducing the charge of the projectile~\cite{Kraft92}. Therefore, $z$
in (\ref{sdcs}) should be replaced by an effective charge $z_{eff}$
which decreases with decreasing energy making the cross section
effectively smaller. In this work, we use $z_{eff}$ given by
Barkas~\cite{Barkas63} which represents the average ion charge at
a given ion velocity $\beta c$:
\begin{equation}
z_{eff}=z(1-\exp(-125\beta z^{-2/3}))~.
\label{zeff}
\end{equation}
The effective charge given by this expression slowly changes at high
projectile velocity, but rapidly decreases in the vicinity of the
Bragg peak. As a result, the charge transfer does not affect the
position of a Bragg peak, but it may significantly affect its height.

Integration of the singly-differentiated cross section over secondary
electron energy $W$ gives the total cross section of impact ionization
by the ion with the kinetic energy $T$:
\begin{equation}
\sigma (T)=\int^{\infty}_0 \frac{d\sigma (W,T)}{dW}dW~.
\label{eq4}
\end{equation}
\noindent
This quantity is important for all our calculations. Infinity as
the upper limit of the integration in this and other expressions that
follow is not physical, but still appropriate because of the
exponential factor in denominator of~(\ref{sdcs}), which effectively
truncates the integral,~\cite{Rudd92}.  

\begin{figure}
\resizebox{1.0\columnwidth}{!}{%
\includegraphics{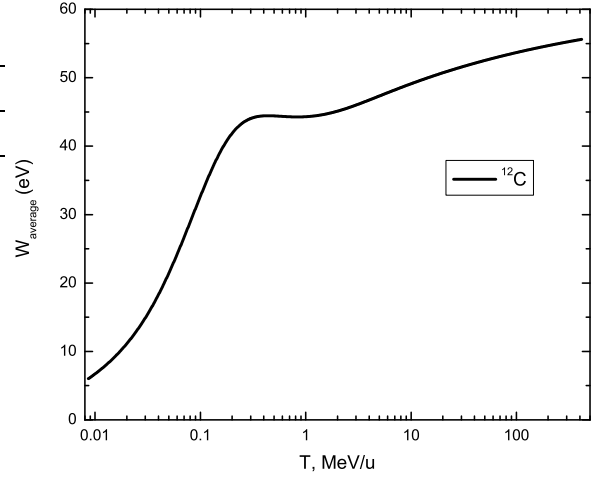}
}
\caption{\label{WaveC} Average energy of secondary electrons produced
as the result of impact ionization as a function of kinetic energy of
$^{12}$C$^{6+}$ ions.}
\end{figure}

The next characteristic that we can obtain from the SDCS is the
average energy of the secondary electrons, which is given by
\begin{eqnarray}
W_{\rm ave}(T)=\frac{1}{\sigma (T)}\int^{\infty}_0 W\frac{d\sigma
(W,T)}{dW}dW~.
\label{eq5}
\end{eqnarray}
\noindent The result of this integration is shown in Fig.~\ref{WaveC}.
It is worth noting that on the average, the secondary electrons are
not very energetic, about 50-60~eV, and the average energy levels out
as the energy of the projectile increases. This fact is important
for estimating the probability of avalanches initiated by the first
generation of emitted electrons.

The stopping cross section, defined as
\begin{eqnarray}
\sigma_{\rm st}=\sum\limits_i \int^{\infty}_0
(W+I_i)\frac{d\sigma_i(W,T)}{dW}dW~,
\label{eq6}
\end{eqnarray}
\noindent gives the average energy lost by a projectile in a single
collision, which can be further translated into energy loss within an
ion's trajectory segment, $d x$:
\begin{eqnarray}
\frac{dT}{dx}=-n\sigma_{st}(T)~.
\label{eq7}
\end{eqnarray}
\noindent This quantity is known as the linear energy transfer (LET),
one of the most important quantities in radiation biology.

\begin{figure}
\vspace{0.5cm}
\resizebox{0.90\columnwidth}{!}{
\includegraphics{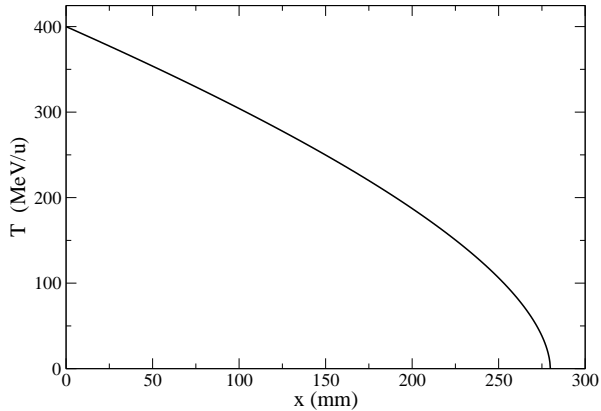}
}
\caption{\label{fig4a} Energy of $^{12}$C$^{6+}$ ions as a function of
the penetration depth in water. The initial ion energy is 400 MeV per
nucleon.  }
\end{figure}
\noindent  
The LET found from Eq. (\ref{eq7}) is a function of the kinetic energy
of the ion rather than the ion's position along the path in the
medium.  The dependence of LET (and other quantities) on this
position, however, is more suitable for cancer therapy applications.
Integrating inverse LET, given by (\ref{eq7}), we get
\begin{eqnarray}
x(T)=\int^{T_{0}}_T \frac{dT'}{|dT'/dx|}~,
\label{eq8}
\end{eqnarray}
\noindent where $T_0$ is the initial energy of the projectile. We
obtain the correspondence between the position of the ion along the
path and its energy.  The dependence $x(T)$ for carbon ions is shown
in Fig.~\ref{fig4a}.

\begin{figure*}
\resizebox{1.90\columnwidth}{!}{
\includegraphics{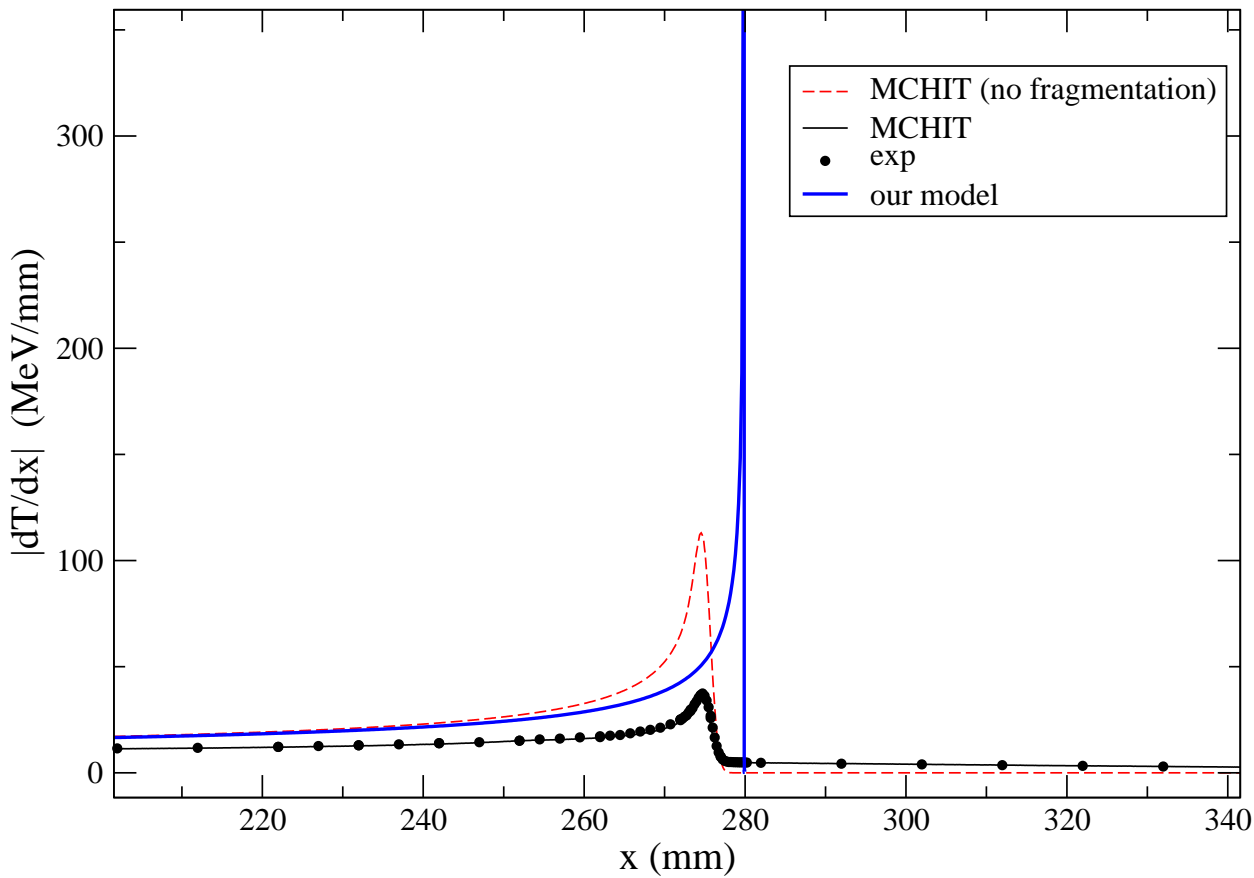}
}
\caption{\label{fig4} Linear energy transfer of 400 MeV/u
$^{12}$C$^{6+}$ ions as a function of penetration depth in water. The
MCHIT results with and without nuclear fragmentation are shown for
comparison. Points - experimental data~\cite{Schardt} }
\end{figure*}

This figure helps to remap all quantities of interest so that they
depend on $x$ rather than on $T$. The depth dependence of the
average LET as a function of $x$ is shown in Fig.~\ref{fig4}.  In this
figure, our calculated LET is compared with the predictions of the
MCHIT model~\cite{Mishustin05,Mishustin06,Mishustin07} based on the
GEANT4 toolkit~\cite{Agostinelli03,Allison06}. The Monte Carlo
calculations were performed both with and without taking into account
nuclear fragmentation reactions.  Experimental data by Schardt {\em et
al.}~\cite{Schardt} are also shown for comparison.
\noindent 
Our calculated position of the Bragg peak is about 5~mm (2\%) deeper
than predicted by simulations and observed by the experiment.
This discrepancy arises from neglecting the projectile energy loss due
to non-ionization processes when the energy is spent for the
excitation of neutral atoms without ejection of electrons. Indeed, it
is estimated~\cite{Kraft92} as 5-15\% of the total projectile energy
loss.
   
\begin{figure*}
\resizebox{1.90\columnwidth}{!}{
\includegraphics{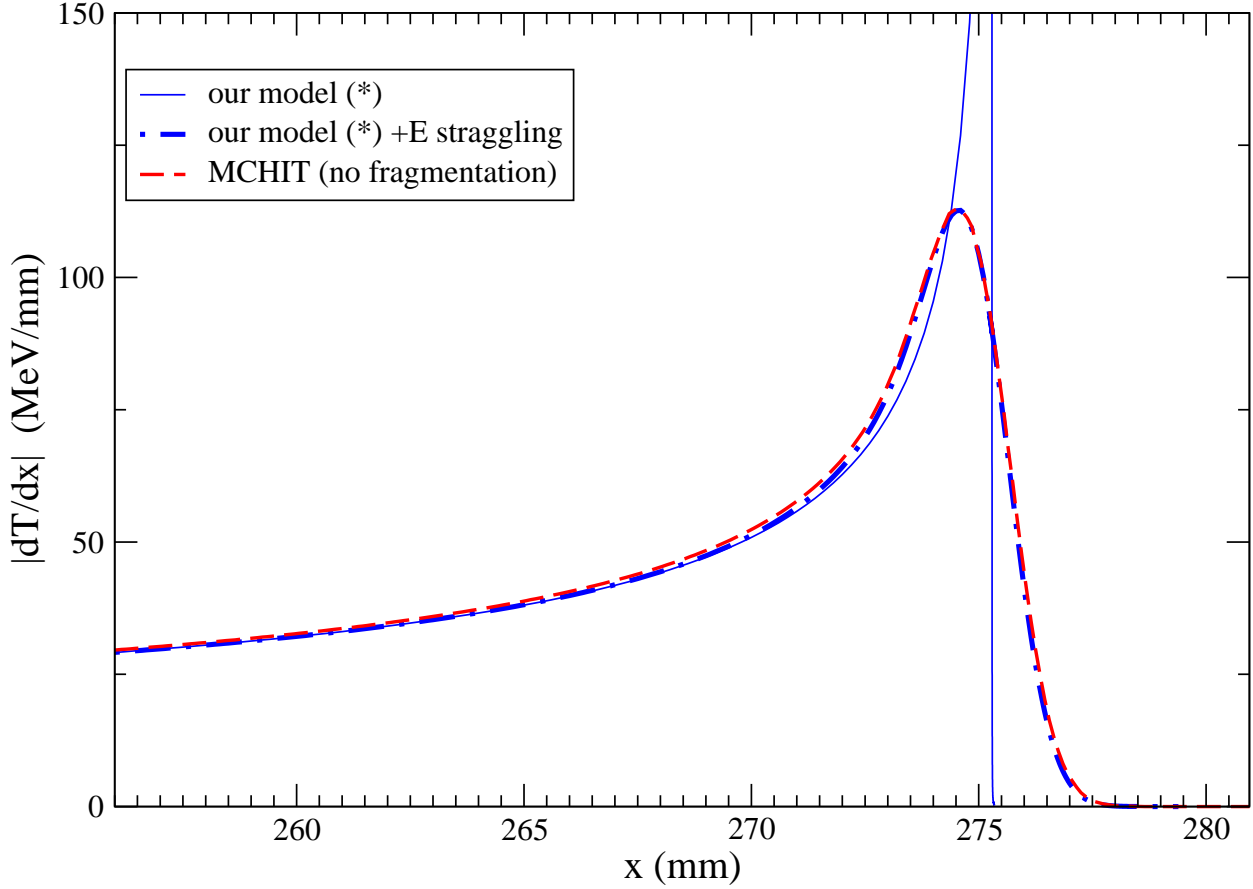}
}
\caption{\label{fig4c} Linear energy transfer of 400 MeV/u
$^{12}$C$^{6+}$ ions as a function of penetration depth in water
with and without straggling compared with the MCHIT results without
nuclear fragmentation. ($^\ast$)Our results are shifted by 4.6 mm for
convenience in comparing the peak shapes.  }  
\end{figure*}

So far our consideration neglects the fact that the projectile energy
loss is a stochastic process, and $|dT/dx|$ was calculated as the
average LET. Therefore, we introduced the energy-loss straggling which
essentially improves the agreement with the MCHIT results obtained
taking into account the stochastics of energy-loss process. 
In calculating the energy-loss straggling, we followed the
prescription of a semi-analytical model described in
Ref.~\cite{Kundrat07},
\begin{eqnarray}
\lefteqn{\left\langle \frac{dT}{dx}(x)\right\rangle
=}\\ \nonumber
&&\frac{1}{\sigma_{str}\sqrt{2\pi}}\int_{0}^{x_0}
\frac{dT}{dx}(x')\exp (-\frac{(x'-x)^{2}}{2\sigma_{str}^{2}})dx'~,  
\label{eqstrg}
\end{eqnarray}
where $x_0$ is a maximum penetration depth of the projectile, and
$\sigma_{str}=0.8$~mm is the longitudinal-straggling standard
deviation computed by Hollmark {\em et al.}~\cite{Hollmark04}, for a
carbon ion of that range of energy.
The
results are presented in Fig.~\ref{fig4c}.  In order to facilitate the
comparison of our analytical model with MCHIT, in Fig.~\ref{fig4c},
MCHIT results are presented without taking into account nuclear
fragmentation reactions, and our data are plotted with a proper shift
in $x$, accounting for the difference between the peak positions.
This allows us to consider separately the 
effects of electromagnetic interactions of beam nuclei.

As confirmed by MCHIT simulations~\cite{Pshenichnov08}, nuclear
fragmentation reactions become important for heavy-nuclei beams and
deeply-located tumors. For example, both experimental
data~\cite{Schardt} and MCHIT calculations~\cite{Pshenichnov07}
indicate that more than 40\% of primary 200 MeV/u $^{12}$C$^{6+}$
nuclei undergo fragmentation before they reach the Bragg peak
position, and this fraction exceeds 70\% for 400 MeV/u $^{12}$C$^{6+}$
beam.  
 
What are the effects of nuclear reactions? First, the beam is
attenuated since ions are leaving the beam. Second, new projectiles are formed
and they may have different penetration depths than the original
carbon ions. Most important, protons and $\alpha$-particles, which
dominate the secondary products of such reactions~\cite{Schardt} have
much longer penetration depths at the same energy (per nucleon). This results in a
substantial tail after the Bragg peak shown in Fig.~\ref{fig4}.

\begin{figure}
\resizebox{1.0\columnwidth}{!}{
\includegraphics{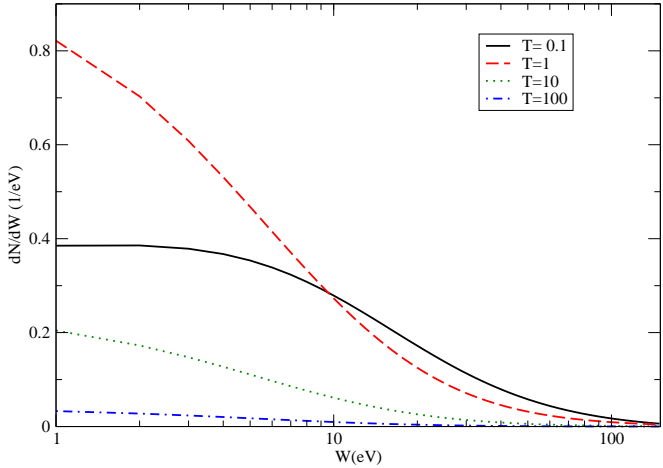}
}
\caption{\label{fig4x} Number of secondary electrons with kinetic
energy $W$, produced by a single ion in a 1-nm interval vs. kinetic
energy of the secondary electrons, for different ion's energies $T$(MeV/u)}
\end{figure}

The next quantity of interest is the number of secondary electrons
with kinetic energy $W$, produced by a single ion on the interval
$\Delta x$ at the depth $x$ corresponding to a certain kinetic energy
$T$ of the ion. This quantity is a product of the remapping of
Eq.~(\ref{eq1}) integrated over solid angle; it can be written as
follows:
\begin{eqnarray}
 \frac{dN(W,x)}{dW}= n\Delta x \frac{d\sigma (W,x)}{dW}~.
\label{eq2}
\end{eqnarray}
Four curves shown in Fig.~\ref{fig4x} show this dependence at
different energies or depths of the ion.
\noindent This number, however, represents only the secondary
electrons produced by the projectile. In order to determine the total
number of electrons in the avalanche, one has to analyze the ionizing
capabilities of these secondary electrons.

\begin{figure}
\resizebox{1.0\columnwidth}{!}{%
\includegraphics{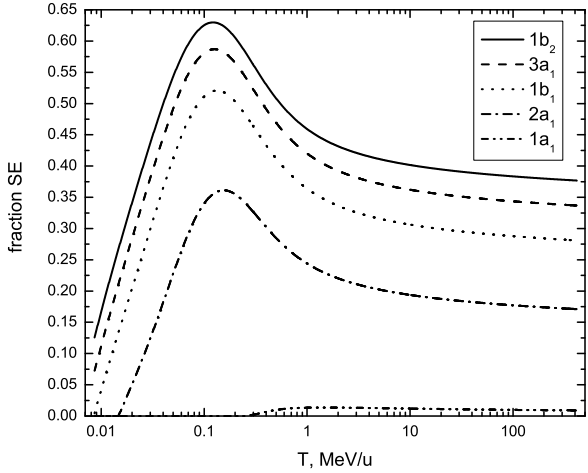}
}
\caption{\label{fig5} Fraction of secondary electrons produced by
$^{12}$C$^{6+}$ capable of further ionization. Different lines indicate
different electronic shells of the water molecule.}
\end{figure}

To address this issue, we calculate the fraction of secondary
electrons with energies higher than the threshold of ionization
produced in a single collision.
\begin{eqnarray}
\phi_i=\frac{1}{\sigma (T)}\int^{\infty}_{I_i} \frac{d\sigma
(W',T)}{dW'}dW'~.
\label{eq9}
\end{eqnarray}
\noindent The dependencies for different electronic shells are shown
in Fig.~\ref{fig5}.
These results are somewhat surprising since they indicate that the
fraction of secondary electrons capable of further ionization
decreases with increasing energy of the ion. This is probably a result
of averaging; even though higher-energetic secondary electrons are
produced, there are many more lower-energy ones (see Fig.~\ref{cs3d}).
The fraction of secondary electrons capable of further ionization is
substantial for secondary electrons produced by the projectile, even
at relatively low projectile energies; however, secondary electrons
produced by {\em secondary} electrons are even less energetic, so that
further ionizations are much less probable.  This is supported by
calculations using Rudd's model for incident electrons. However, it
can also be explained by a simple estimate: Fig.~\ref{WaveC} indicates
that the average energy of secondary electrons in the vicinity of the
Bragg peak is about 45~eV. The maximum average energy that can be
transferred to the next generation secondary electron is just
$(45-I_i)/2$, which is just over 12~eV for the outermost electron, an
energy barely enough to cause further ionization.  Therefore, we
conclude that there is no avalanche of secondary electrons and that
Eq. (\ref{eq2}) gives a reasonable estimate of the number of secondary
electrons, within a factor of 2.
Thus, we have analyzed the track of a single ion: we obtained the LET
(depicting the Bragg peak, its shape and position), we obtained the
energy spectrum of the secondary electrons, and we estimated the
avalanche effects related with ionization caused by secondary
electrons.

\section{Direct consequences of ionization and DNA damage}
\label{estimates}

In this section we estimate three components that contribute to the
DNA damage: an electron plasma, free radicals produced as a result of
ionization of water molecules, and heating of the medium caused by ion
passage.

\subsection{Electron plasma and free radicals}
\label{plasma}

Let us estimate the number density of the electron plasma produced by
an ion. In Section~\ref{ionpass}, we found the secondary electron
energy spectrum and estimated the effect of further ionization. In
order to deduce the density of the electron plasma, we have to estimate
the volume occupied by these electrons. The crucial quantity for this
estimate is the penetration length of the secondary electrons caused by
the projectile. In principle, this estimate can be done in the same
fashion as it was done for the ions in the previous section; however,
the energies of the secondary electrons are such that Rudd's
analysis may not be reliable~\cite{Rudd92}. Therefore, we will try to
rely on other available sources.

Direct experimental observations of penetration depths of electrons in
liquid water are scarce.  There are values recommended by the
International Commission on Radiation Units and Measurements for 10 --
1000 keV electrons \cite{ICRU,ESTAR}, and there are data for
thermalization lengths of ``subexcitation electrons'' for lower
energies, 0 -- 4 eV \cite{Konovalov1988}.  However, there is a gap in
experimental data for electron energies, which are the most
interesting from the viewpoint of 
cancer therapy applications.

The average energy of secondary electrons in the Bragg peak area
($T\sim 0.3$ MeV per nucleon) is about 45~eV (see Fig.~\ref{WaveC}).
Even though there are no direct experimental values for the
penetration lengths of electrons in this energy range, interpolation
of existing experimental data points~\cite{Watt96} and theoretical
predictions based on Monte-Carlo simulations~\cite{Meesungnoen02}
suggest that the penetration lengths of such electrons are of the
order of 10 nm or less.  Comparison of these lengths and typical
distances between the ions in the beam leads us to conclude that under
realistic conditions ion tracks do not interact. 
That is, every ion acts independently and the total biological effect
equals the sum of the damages from individual ions, and therefore the
collective action of ions in the beam can be neglected.

In order to get realistic order of magnitude estimates, we
assume that the energy deposited by an ion is thermalized within a
tube of a radius $r\sim 10$ nm.

The total ionization cross section for carbon ions $^{12}$C$^{6+}$ of
kinetic energy $T\sim 0.3$ MeV per nucleon is about $\sigma \approx
40\cdot 10^{-16} ~ {\rm cm}^{2} = 0.4 ~ {\rm nm}^{2}$.  The majority
of secondary electrons are produced in the following process:
\begin{equation}
\label{ion}
^{12}{\rm C}^{6+} + {\rm H}_2{\rm O} \rightarrow ^{12}{\rm C}^{6+} + {\rm H}_2{\rm
O}^{+} + e^-.
\end{equation}
\noindent 
The number of electrons produced per one nm of the ion's trajectory is
then $n \sigma = 33~ {\rm nm}^{-3}\times 0.4~ {\rm nm}^{2} \approx 13
~{\rm nm}^{-1}$.  This quantity is easily translated into the number
density of the electron plasma,
\begin{equation}
n_e = \frac{n \sigma}{\pi r^2} \approx \frac{13~ {\rm nm}^{-1}}{\pi
100~ {\rm nm}^2} \approx 0.045~ {\rm nm}^{-3} = 0.45\cdot10^{20} ~{\rm
cm}^{-3}.
\end{equation}

Both secondary electrons and ionized water molecules react to form
free radicals
\begin{equation}
\label{reaction1}
{\rm H}_2{\rm O}^+ \rightarrow {\rm H}^+ + {\rm OH}\cdot
\end{equation}
\begin{equation}
\label{reaction2}
{\rm H}_2{\rm O} + e^- \rightarrow {\rm H}\cdot + ~ {\rm OH}^-~.
\end{equation} 
This means that the induced concentrations of free radicals can be of
the same order as the concentration of the electron plasma.

OH radicals play an important role in damaging DNA. They
interact with nucleobases damaging them; moreover, ``secondary'' free
radicals may result from these interactions~\cite{Nikjoo06}. Further 
analysis is needed in order to obtain quantitative estimates of the
yields of DSB's and other types of irreparable DNA damage due to
these effects.  

\subsection{Dissociative recombination by low energy electrons}

Traditionally, it had been believed that non-thermalized secondary
electrons can not cause any significant genotoxic damage, unless they
are already solvated in water and can participate in chemical
reactions of the type (\ref{reaction2}) or unless they have enough
kinetic energy. The singly differentiated cross section, however,
falls rapidly with energy of the ejected electron and therefore
energetic secondary electrons are not abundant, cf. Figs.~\ref{cs3d}
and \ref{WaveC}.

In 2000, Sanche {\em et al.}~\cite{Sanche2000} 
 argued that 1- to 20-eV secondary electrons
``induce substantial yields of single- and double-strand breaks in
DNA, which are caused by rapid decays of transient molecular
resonances localized on the DNA's basic
 components''~\cite{Sanche2000}. In their 
experiments, a monoenergetic electron beam at various incident energies
was used to irradiate solid DNA samples. They observed that one
electron out of 1000 -- 2000 (depending on the electron energy)
induced a single strand break and one electron out of 5000 -- 10000
induced a DSB of a DNA molecule~\cite{Sanche2003}.

This type of experiment, however, can not be directly related to the
problem of DNA damage induced by ions.  The difference is in the
density of free electrons.  In Sanche's experiments the electron
density was of the order of $10^4~{\rm cm}^{-3}$, 16 orders of
magnitude lower than is created by an ion in the Bragg's peak area, as
it is estimated in the previous section.

One can nevertheless estimate the average number of electrons created
in a volume occupied by one convolution of the DNA molecule.  The DNA
radius is about 1 nm, while the convolution (linear) length is about 4
nm, so that the volume of interest is about $V_{\rm DNA} = 10^{-20}~
{\rm cm}^{3}$.  Assuming that at least two electrons are needed in the
volume of one DNA convolution to produce a double strand break, one
can estimate that the required electron density
\begin{equation}
n_e^{\rm crit} = \frac{2}{V_{\rm DNA}} \approx 2\cdot10^{20}~ {\rm
cm}^{-3}.
\end{equation}
\noindent It is remarkable that such a density is of the same order of magnitude of the one that has been predicted by our simple estimates in the previous section.

\subsection{Local heating}

Finally, let us estimate the local heating produced in a tube of 10 nm
radius by a carbon ion of 0.3 MeV per nucleon. The simplest way to do
that is to use the thermodynamic equation
\begin{equation}
Q = \mu c \Delta T
\end{equation}
\noindent relating the heat transferred, $Q$, to a system with the
system's mass $\mu$, specific heat capacity $c$ (4.2 J/g K for water)
and the increase of system's temperature $\Delta T$. $\Delta T$ can
then be found from the linear energy transfer $Q/\Delta x$ (typical
value for $^{12}$C ions is 100 MeV/mm = $1.6 ~ 10^{10}$J/cm in the
Bragg peak region), the tube's radius $r$ and the density of water
$\rho = 1~ {\rm g}/{\rm cm}^{-3}$:
\begin{equation}
Q = \rho \pi r^2 \Delta x c \Delta T,
\end{equation}
\begin{equation}
\Delta T = \frac{Q/\Delta x}{\rho \pi r^2 c}.
\end{equation}
\noindent Substituting the numerical values, one obtains $\Delta T
\sim 10 K$ for $r=10$ nm and for the lower
estimate of the tube's radius, $r=3$ nm, we find $\Delta T \sim 100 K$.

This estimate shows that the local heating within the electron
thermalization radius along the ion track can be quite
substantial. The melting temperature of a DNA molecule is about
85$^o$ C. 
 Therefore, in our opinion, there is a need for a
more thorough study of the local heating mechanism which would include
investigation of such problems as the detailed description of the
thermalization of secondary electrons in water, heat transfer in
water, and modeling of DNA dynamics under local heating.  To the best
of our knowledge, this local heating mechanism has not yet received
any attention.

\section{Conclusions}
We have presented an approach to modeling ion-beam therapy
having considered the effects initiated 
by an energetic projectile such
as a carbon ion incident on biological tissue (represented by liquid
water). We analyzed passage of the ion through this medium taking into
account the main processes that cause energy loss by the
projectile. We succeeded in making quantitative predictions of effects
such as the energy spectrum and abundance of secondary
electrons and local heating caused by the projectile. Then we built
our estimates of DNA damage on these predictions. In principle, the
final estimates may be related to the energy deposition by the
projectiles, but the microscopic analysis of the whole scenario is
vitally important in order to provide more accurate predictions that
may eventually contribute to the protocol of cancer therapy. Then they
can be formulated into the language of dosages, energies, radiation
rates, {\em etc.} Before that, more research is needed in
understanding the mechanisms of DNA damage on a microscopic level and
relating these mechanisms to the physical parameters of secondary electrons and
free radicals produced by the ion beam. 

\section*{Acknowledgments}
This work is partially supported by the European Commission within the
Network of Excellence project EXCELL and by the EU project PECU. E.S. is
grateful to J.S. Payson for fruitful discussions and
to FIAS for hospitality and support.

\end{document}